\documentclass[prb,twocolumn,preprintnumbers,superscriptaddress]{revtex4-2}

\usepackage{amsmath}
\usepackage{xcolor}
\usepackage{graphicx}
\usepackage{dcolumn}
\usepackage{bm}
\usepackage{hyperref}
\usepackage[english]{babel}
\DeclareMathOperator{\e}{e}

\begin{document}


\title{
	Coherent superposition of emitted and resonantly scattered photons from a two-level system in a cavity driven by an even-$\pi$ pulse
}

\author{I.V.~Krainov}
\email{igor.kraynov@mail.ru} 
\affiliation{Ioffe Institute, 194021 St.~Petersburg, Russia} 
\author{A.I.~Galimov}
\affiliation{Ioffe Institute, 194021 St.~Petersburg, Russia} 
\author{Yu.M.~Serov}
\affiliation{Ioffe Institute, 194021 St.~Petersburg, Russia} 
\author{A.A.~Toropov}
\affiliation{Ioffe Institute, 194021 St.~Petersburg, Russia} 
\author{T.V.~Shubina}
\affiliation{Ioffe Institute, 194021 St.~Petersburg, Russia} 

\date{\today}

\begin{abstract}
We examine the nature and quantum satistics of a multiphoton bunch generated under resonant excitation by even-$\pi$ pulses  of a  charged quantum dot in a cavity. Experiments with Rabi oscillations in time-resolved photoluminescence, varying pulse duration, and cavity mode detuning show that this multiphoton state is a coherent superposition of emitted photons and resonantly scattered laser photons. The developed model incorporates both resonant scattering and re-excitation mechanisms, allowing us to describe the quantum statistics of the multiphoton state, which is controlled by the coupling between the quantum dot and optical modes in the cavity.

\end{abstract} 

\maketitle

A two-level system - a semiconductor quantum dot (QD) embedded in a microresonator - is a promising resource for quantum technologies, capable of producing both single-photon Fock states \cite{Senellart2017, Rakhlin2023} and multiphoton entangled states \cite{Tiurev2022, Huet2025}. The system provides a flexible platform for studying the fundamentals of cavity quantum electrodynamics \cite{Jaynes1963, Moelbjerg_2012}, as well as various states of classical and quantum light. It is noteworthy that the clear boundary in their application is blurred, and different superpositions may be of interest. For example, entangled states reminiscent of Schr\"odinger's cat \cite{Ourjoumtsev2007}, obtained by the interference of classical and quantum light \cite{Afek2010}, occur useful for quantum key distribution \cite{Hwang2003}.

The optical response of a QD-cavity system strongly depends on the excitation type. A well-established method for controlling that is resonant excitation followed by cross-polarization filtering of the laser light \cite{Muller2007, Galimov2021}. It has been shown theoretically that with increasing pump power, a quantum emitter in a cavity can emit a bunch of N photons, where N can be any integer \cite{Munoz2014}. 
Various non-classical multiphoton states have been studied experimentally, both with and without a cavity, such as two-photon bunches \cite{Fischer2017}, photon number superpositions \cite{Loredo2019}, photon number entangled states \cite{Wein2022}, and dynamic resonance fluorescence with side peaks \cite{Boos_2024,Liu2024}.

The emission character of a QD differs significantly at the two poles of excitation power, corresponding to a pulse area with odd-$\pi$ and even-$\pi$ multiplicity. At odd $\pi$, the second-order correlation function $g^{(2)}(0)$, which characterizes photon statistics, tends to zero, indicating the emission of single photons. At even $\pi$, multiphoton bunches are created, the $g^{(2)}(0)$ function of which can vary from sub-Poissonian to super-Poissonian depending on the pulse duration and the presence of a cavity.

To explain the formation of  multiphoton bunches in a QD with a trion, Fischer et al. proposed a re-excitation model \cite{Fischer2017,Fischer_2018}, which is based on the assumption of a non-zero probability that the first part of the pulse will excite the QD so that it will emit a photon, and the remaining part of the pulse will be able to re-excite the QD, which will emit another photon during the trion lifetime.
Re-excitation can certainly promote the formation of a multiphoton state. However, the proposed model describes the interaction of the pump pulse with a QD without a cavity, as was the case in their experiments, whereas in many cases \cite{Giesz_2016, Loredo2019, Liu2024, Giorgino_2025} the pulse interacts with the QD-cavity system, which dramatically changes the optical response \cite{Senellart2008}.

To demonstrate the cavity effect, Fig. 1 shows the dependencies of $g^{(2)}(0)$ on the pulse duration, measured in systems with and without a cavity. The data we measured using a QD in the cavity are in good agreement with previously published data for similar systems \cite{Loredo2019, Giorgino_2025}. However, they  differ sharply from the data for a cavity-free QD, taken from \cite{Fischer2017}. Successfully modeling the data obtained with short-pulse excitation using the re-excitation model is problematic. This highlights the need for an extended approach that takes cavity effects into account.

It should also be noted that a two-level system can not only emit photons but also scatter light resonantly, which is not taken into account in the re-excitation model \cite{Fischer2017,Fischer_2018}. 
The scattering can occur under weak excitation in the Heitler regime \cite{Hanschke2020, Masters2023} and under strong excitation in the Mollow regime, initially under continuous excitation \cite{Mollow1969}.The term "resonance fluorescence" instead of "resonance scattering" used in \cite{Mollow1969} was proposed by Moelbjerg et al. in 2012 for pulse excitation \cite{Moelbjerg_2012}. When studying this dynamic process, which leads to the appearance of side peaks around the main peak \cite{Boos_2024, Liu2024}, the emission spectrum was characterized by a first-order correlation function, which does not provide information about the nature of the emitted light, namely, the origin of the photons and their quantum statistics.

In this Letter, we explore the nature of the multiphoton bunch generated by even-$\pi$ pulses in a charged QD-cavity system and show that it is a coherent superposition of emitted and scattered photons controlled by QD-cavity coupling. The composition is characterized by the second order correlation function $g^{(2)}(0)$. An exact theoretical model is presented that does not require (but does not exclude) spontaneous photon emission during the excitation pulse. We describe criteria for resonant light scattering to be significant and propose a simple "box" model for simulating experimental $g^{(2)}(0)$ dependences and Rabi oscillations in multiphoton emission.

\begin{figure}[t] 
\includegraphics[width=0.8\linewidth] {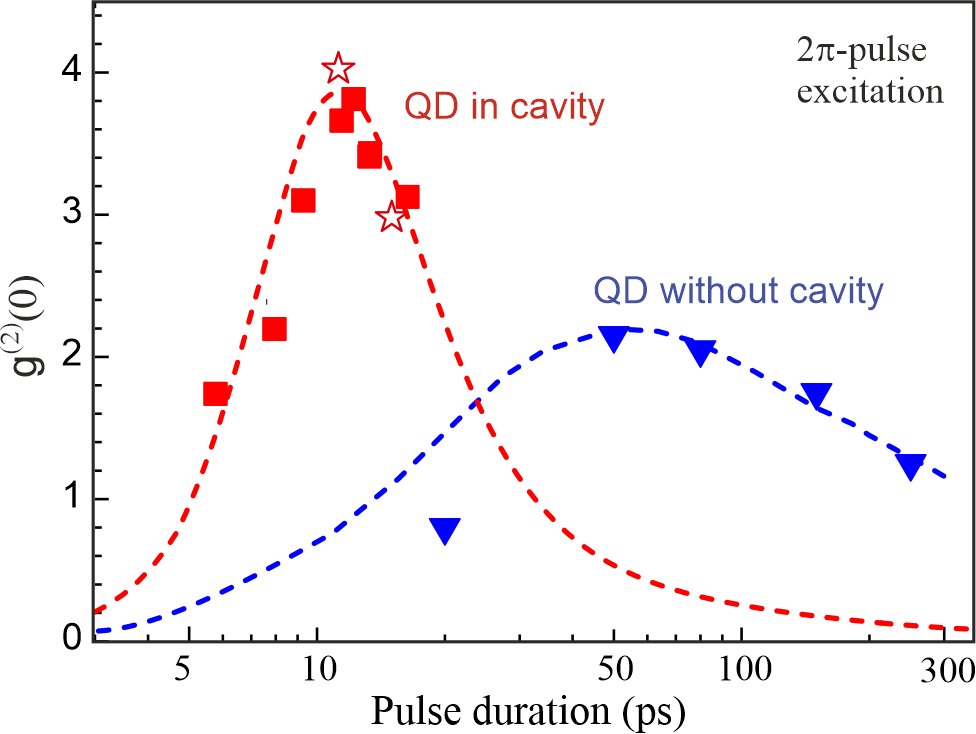} 
\centering 
\caption{Cavity effect on the dependence of $g^{(2)}(0)$ on the 2$\pi$-pulse duration.
For a QD in a resonator (red), the squares represent our measurements; the asterisks denote the  values from \cite{Loredo2019} and \cite{Giorgino_2025} (shown at the peak of the dependency).  Modeling details are in the Supplementary. For a QD without a cavity (blue), the experimental data and their simulation are taken from \cite{Fischer2017}.}
\label{fig1}
\end{figure}

We consider a charged QD with a hole placed in a symmetric microresonator. The polarization of the optical modes $i = \text{H, V},\pm$ can be linear and circular, $\pm\sigma$. During resonant excitation, the energy of the fundamental cavity mode $\hbar \omega_0$ coincides with that of the trion transition $E_t$ (other modes are ignored). The rotational wave approximation is used. Under these conditions, the interaction of the system with the optical mode is described by the Hamiltonian:
\begin{gather}
\label{hamH0}
    \hat{H}_0 = g\, \hat{b}^\dag_- \hat{a}^\dag_+ \hat{d}_+ + g^*\, \hat{b}_- \hat{a}_+ \hat{d}^\dag_+ + 
    g^*\, \hat{b}^\dag_+ \hat{a}^\dag_- \hat{d}_- + g\, \hat{b}_+ \hat{a}_- \hat{d}^\dag_-,
\end{gather}
where  the mode creation operator $\hat{b}^\dagger_\pm = (\hat{b}^\dagger_H \pm i\hat{b}^\dagger_V)/\sqrt{2}$, the hole creation operator is $\hat{a}^\dagger_\pm$, where $\pm$ denotes the heavy hole state $\left|\pm3/2 \right\rangle$, the trion creation operator is $\hat{d}^\dagger_\pm$, where $\pm$ denotes a positively charged trion with electron spin $\left|\pm1/2 \right\rangle$, and $g$ is the QD–cavity coupling parameter.

To correctly describe the dynamics of the entire system, in addition to Eq. (\ref{hamH0}), the  escape  of photons into the environment must be taken into account \cite{Manzano2020}:
\begin{gather}
\label{linbEq}
    \partial_t \hat{\rho} = -\frac{i}{\hbar} \left[ \hat{H}_0, \hat{\rho} \right] + 
    \sum_i \gamma_i \left( \hat{V}_i \hat{\rho} \hat{V}^\dag_i - 
\frac{1}{2}\left\{ \hat{V}^\dag_i \hat{V}_i , \hat{\rho} \right\} \right)\equiv \hat{\mathcal{L}} (\hat{\rho}), 
\end{gather} 
where $\{\cdot, \cdot \}$ is the anticommutator, $\gamma_i$ and $ \hat{V}_i$ are the damping rate and the jump operator in the Lindbladian. For resonant scattering, we  will consider only the photon yield from the cavity mode, assuming that its rate is equal to the mode width $\Gamma$, and the jump operator $\hat{V}_{phot} = \hat{b}_\pm$. 

Consider resonant excitation by a linearly $H$-polarized pulse. Since the pulse contains a large number of photons, we can replace the $H$-mode operators with a classical field through the Rabi frequency $g\hat{b}_H \rightarrow \Omega_R(t)$.
Before the arrival of the pump pulse, the hole is in an incoherent superposition of states with spin $\left|\pm3/2 \right\rangle$ with equal probability $\hat{\rho}_0 = (\hat{a}^\dag_+ \left|0 \right\rangle \left\langle 0 \right| \hat{a}_+ + \hat{a}^\dag_- \left|0 \right\rangle \left\langle 0 \right| \hat{a}_- ) / 2$.

The second-order correlation function at zero delay time is given by the expression:
\begin{gather}
    g^{(2)}(0) = \frac{P_2}{(P_1 + P_2)^2},
\end{gather}
where the probabilities of detecting one or two photons, $P_1$ or $P_2$, are expressed through the evolution operator $\exp(\hat{\mathcal{L}}t)$ in equation (\ref{linbEq}) as follows:
\begin{gather}
   P_1 = \Gamma \int_{0}^{\infty}dt\, \text{Tr} \left[
    \hat{b}_V \e^{\hat{\mathcal{L}}t} \left(\hat{\rho}_0 \right) \hat{b}^\dag_V
    \right], \\
    P_2 = \Gamma^2 \int_{0}^{\infty}dt_2 dt_1 \text{Tr} \left[\hat{b}_V \e^{\hat{\mathcal{L}}(t_2-t_1)} \left( \hat{b}_V \e^{\hat{\mathcal{L}}t_1} \left(\hat{\rho}_0 \right) \hat{b}^\dag_V \right)\hat{b}^\dag_V \right]. 
\end{gather}

\begin{figure*}[t]
\includegraphics[width=0.95\linewidth]{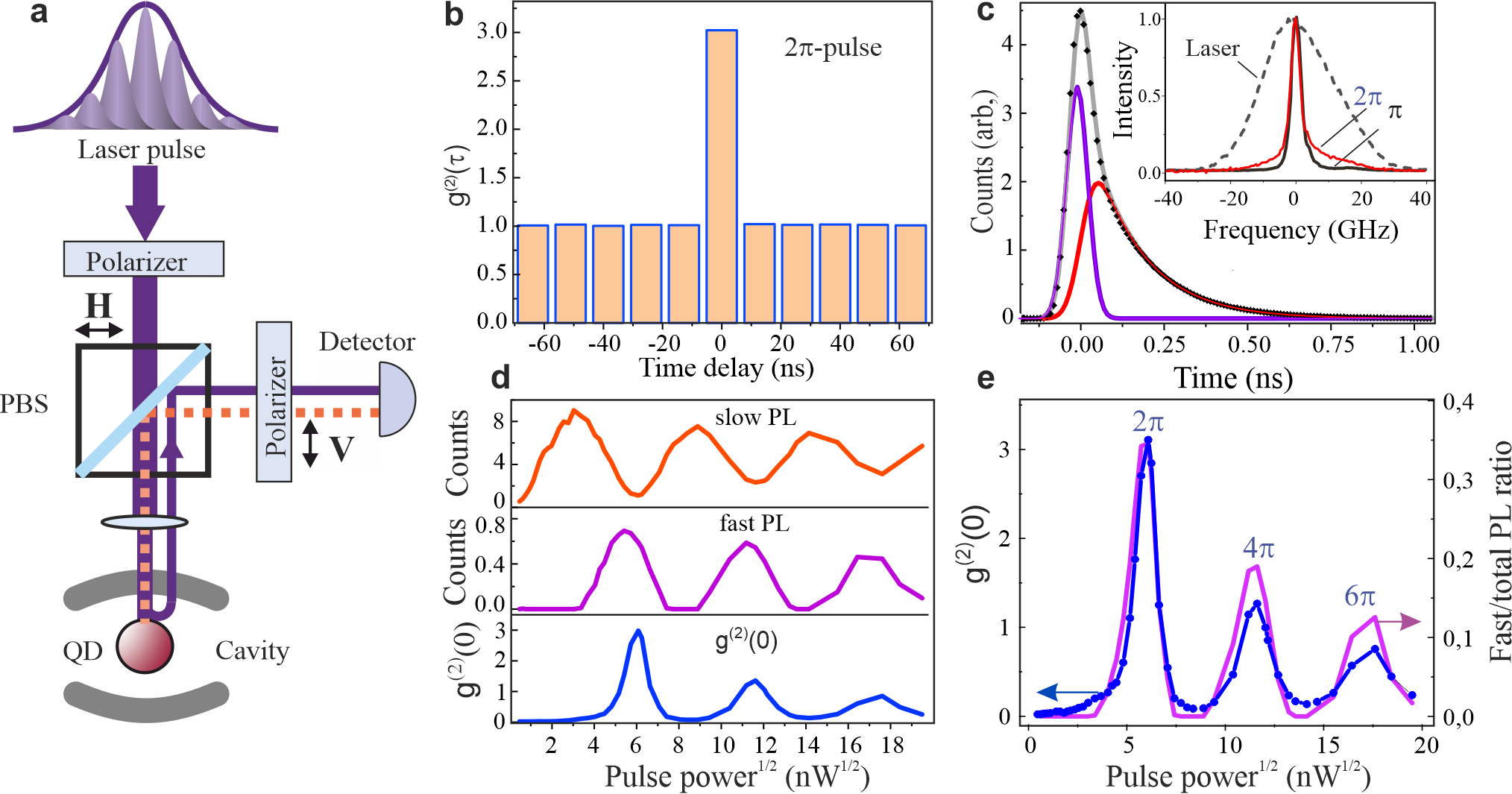}
\centering
\caption{\label{fig1k} 
(a) The InAs/GaAs QD in the microcavity is resonantly excited by a H-polarized 16-ps laser pulse. A cross-polarization scheme with a polarizing beam splitter (PBS) allows only V-polarized light to pass to a single-photon detector.
(b) Second-order correlation function $g^{(2)}(t)$ of QD emission excited by the $2\pi$-pulse. The bin width is equal to the laser pulse repetition period (12.7 ns). (c) Time-resolved PL decay curve for $2\pi$ excitation and its decomposition into a fast scattered light and slow trion emission with a characteristic decay time of 155 ps. The inset shows narrow PL lines excited by $\pi$-pulse (black) and a $2\pi$-pulse (red). The dashed line presents laser pulse spectrum. The zero corresponds to 1350.75 meV. 
(d) Rabi oscillations in the slow PL component, fast PL component, and $g^{(2)}(0)$ as functions of excitation pulse power. The counts are in MHz. 
(e) Strong correlation between the function $g^{(2)}(0)$ and the ratio of the fast PL intensity to the total PL intensity.
}
\end{figure*}

This approach does not ignore the coherent states $\hat{b}^{\dag n}_V \hat{a}^\dag_+ \left|0 \right\rangle, \hat{b}^{\dag n}_V \hat{d}^\dag_+ \left|0 \right\rangle$, which correspond to  the case when the photons of mode $V$ do not leave the cavity immediately, before some changes in the coherent dynamics of the system occur. This case corresponds to a high density of optical states $\Gamma \lesssim \Omega_R$. At a low density of optical states $\Gamma \gg \Omega_R$, as without an optical cavity, these states can be neglected. If we restrict ourselves to only single-photon states $\hat{b}^{\dag}_V \hat{a}^\dag_+ \left|0 \right\rangle, \hat{b}^{\dag}_V \hat{d}^\dag_+ \left|0 \right\rangle$ (higher states with $n>1$ will be excited with less probability), it becomes clear that two processes are possible, namely:
a re-excitation process associated with the transition
\begin{gather}
\label{reExProc}
    \hat{a}^\dag_+ |0\rangle \xrightarrow[\Omega_R]{} 
    \hat{d}^\dag_+ |0\rangle \xrightarrow[g]{} \hat{b}^\dag_V \hat{a}^\dag_+ |0\rangle \xrightarrow[\Gamma]{} \hat{a}^\dag_+ |0\rangle \xrightarrow[\Omega_R]{} \hat{d}^\dag_+ |0\rangle,
\end{gather}
and a scattering process associated with the emergence of two-photon states from the coherent dynamics
\begin{gather}
\label{resScatProc}
    \hat{a}^\dag_+ |0\rangle \xrightarrow[\Omega_R]{} 
    \hat{d}^\dag_+ |0\rangle \xrightarrow[g]{} \hat{b}^\dag_V \hat{a}^\dag_+ |0\rangle \xrightarrow[\Omega_R]{}  \hat{b}^\dag_V \hat{d}^\dag_+ |0\rangle.
\end{gather}
To compare their contributions, we can consider the dynamics in first order in $g/\Gamma$, assuming a rectangular pump pulse of duration $\tau$ and neglecting phonons.
Since the Hamiltonian (\ref{hamH0}) does not mix trion states, for simplicity we can consider only one state $\left|+3/2 \right\rangle$ (for the opposite spin, the situation will be similar). Using the approximation of the non-Hermitian Hamiltonian describing the dynamics (\ref{linbEq}), we obtain the two-photon contributions for the re-excitation (\ref{reExProc}) and scattering (\ref{resScatProc}) processes:
\begin{gather}
\label{p2reEx}
    P_2^{\text{re-ex}} = \Gamma \int_0^\tau dt\, \langle 0| \hat{a}_+ \hat{b}_V \hat{\rho}(t) \hat{b}^\dag_V \hat{a}^\dag_+ |0\rangle \sin^2(\Omega_R (\tau-t)), \\  \nonumber
    P_2^{\text{scat}} = \Gamma \int_0^\tau dt\, \langle 0| \hat{d}_+ \hat{b}_V \hat{\rho}(t) \hat{b}^\dag_V \hat{d}^\dag_+ |0\rangle \cos^2(\Omega_R (\tau-t)) \\ 
    + \langle 0| \hat{d}_+ \hat{b}_V \hat{\rho}(\tau) \hat{b}^\dag_V \hat{d}^\dag_+ |0\rangle,
\label{p2scat}
\end{gather}

In the expression (\ref{p2reEx}), the term $\sin^2(\Omega_R (\tau-t))$ describes the re-excitation of the trion after spontaneous photon emission during the pump pulse. In (\ref{p2scat}), $\cos^2(\Omega_R (\tau-t))$ describes the probability that the trion is  in an excited state  during the pump pulse and after resonant photon scattering. The last term in (\ref{p2scat}) describes the probability that after the end of the excitation pulse the system is in a state with an excited trion and a resonantly scattered photon in the mode. For both expressions (\ref{p2reEx} and \ref{p2scat}), it is assumed that after the pump pulse, according to Eq.(\ref{linbEq}), the excited trion will emit a photon with probability close to unity.

To estimate the contribution of these two mechanisms, we consider the regime of short pump pulses $\Gamma \tau = 1$, assuming that $\Omega_R = \pi \Gamma$ for a pulse $2\pi$, which gives the probabilities:
\begin{gather}
    P_2^{\text{re-ex}} \approx 0.01\, \frac{g^2 }{\Gamma^2}, \quad
    P_2^{\text{scat}} \approx 0.1\, \frac{g^2 }{\Gamma^2}.
\end{gather}
These estimates show that the contribution of resonant scattering is extremely important for short pump pulses. For long pulses, $\Gamma \tau \gg 1$, the contribution of resonant scattering decreases due to the decay of the involved coherent states, while the contribution of re-excitation increases proportionally to $\tau$. The dependencies presented in Fig.~1 are in agreement with  these trends.


To demonstrate the existence of resonance photon scattering in a QD-cavity system, we used a positively charged InAs/GaAs QD embedded in a micropillar resonator.  Resonant excitation was done by a short 16-ps pulse.
The cross-polarization measurement scheme shown in Fig.~2a provides attenuation of the exciting laser light by six orders of magnitude \cite{Galimov2021}. Further details on the sample fabrication and measurements are provided in the Supplementary.

When excited by a $\pi$ pulse, we observed photon antibunching with $g^{(2)}(0) = 0.05 \pm 0.01$, confirming the emission of single photons. When excited by a $2\pi$ pulse,  we recorded bunching with $g^{(2)}(0) = 3.0 \pm 0.3$ (Fig.~2b), as in \cite{Loredo2019} for similar conditions.
Then, we measured time-resolved photoluminescence (PL) decay curves by varying the pulse power up to 6$\pi$. For odd $\pi$, the decay curves are mono-exponential, whereas for even $\pi$, they are bi-exponential. Figure 1c shows the decomposition of the $2\pi$ decay curve into two components, fast and slow, presumably originating from resonant scattering of laser photons and spontaneous emission of the trion.

Figure 2d shows that the slowly decaying PL component has maxima at odd $\pi$  and minima at even $\pi$, as expected for a photon emission probability controlled by the pulse area. In contrast, the fast component exhibits extrema at the opposite $\pi$-multiplicity. With measurements at the same power values, the $g^{(2)}(0)$ maxima coincide with those of the fast component. More precisely, the $g^{(2)}(0)$ oscillations reproduce the ratio of the fast component intensity to the total PL intensity (Fig.~2e). It is noteworthy that at  2$\pi$, the fast component accounts for approximately one-third of the total PL intensity, while the expected contribution from re-excitation (10) is significantly smaller for such short pulses. These facts confirm that multiphoton emission is primarily due to fast photon scattering during the pump pulse and is independent of the system's state after its completion.

The PL lines excited by the $\pi$ and $2\pi$ pulses have the same $\sim$10 $\mu$eV width, and their intensities differ by a factor of six. Both lines have wings, resembling the side peaks in the dynamical resonance fluorescence \cite{Boos_2024, Liu2024}. Their integrated intensity in the $2\pi$ spectrum is comparable to that of the narrow central peak.  
With strong cross-polarization filtering, this contribution can be enhanced by rotating the original light polarization in the QD-cavity system. Indeed, a single spin in a charged QD without a cavity can provide a rotation of only a thousandth of a degree. In contrast, the spin-photon interaction is significantly enhanced when a charged QD is coupled to a cavity, resulting in a polarization rotation of several degrees \cite{Androvitsaneas2916, Arnold2015, Chan2023, Mehdi2024}.

To demonstrate the influence of the coupling of a QD to a cavity on the statistics of multiphoton bunches, we conducted an experiment on the detuning of the QD emission line relative to the cavity mode, achieved by their different temperature dependencies. In this experiment, a 2$\pi$ laser pulse tracked the trion transition energy. With detuning, we observed a decrease in the Purcell factor, manifested by an increase in the trion PL decay time (Fig.~3a). Resonant photon scattering is also decreased due to the suppression of the nonlinear rotation effect. As a result, the photon statistics changes from super-Poissonian to near-Poissonian (Fig.~3b). Ultimately, laser photons randomly scattered outside the optical mode will dominate.


\begin{figure}[t]
\includegraphics[width=1\linewidth]{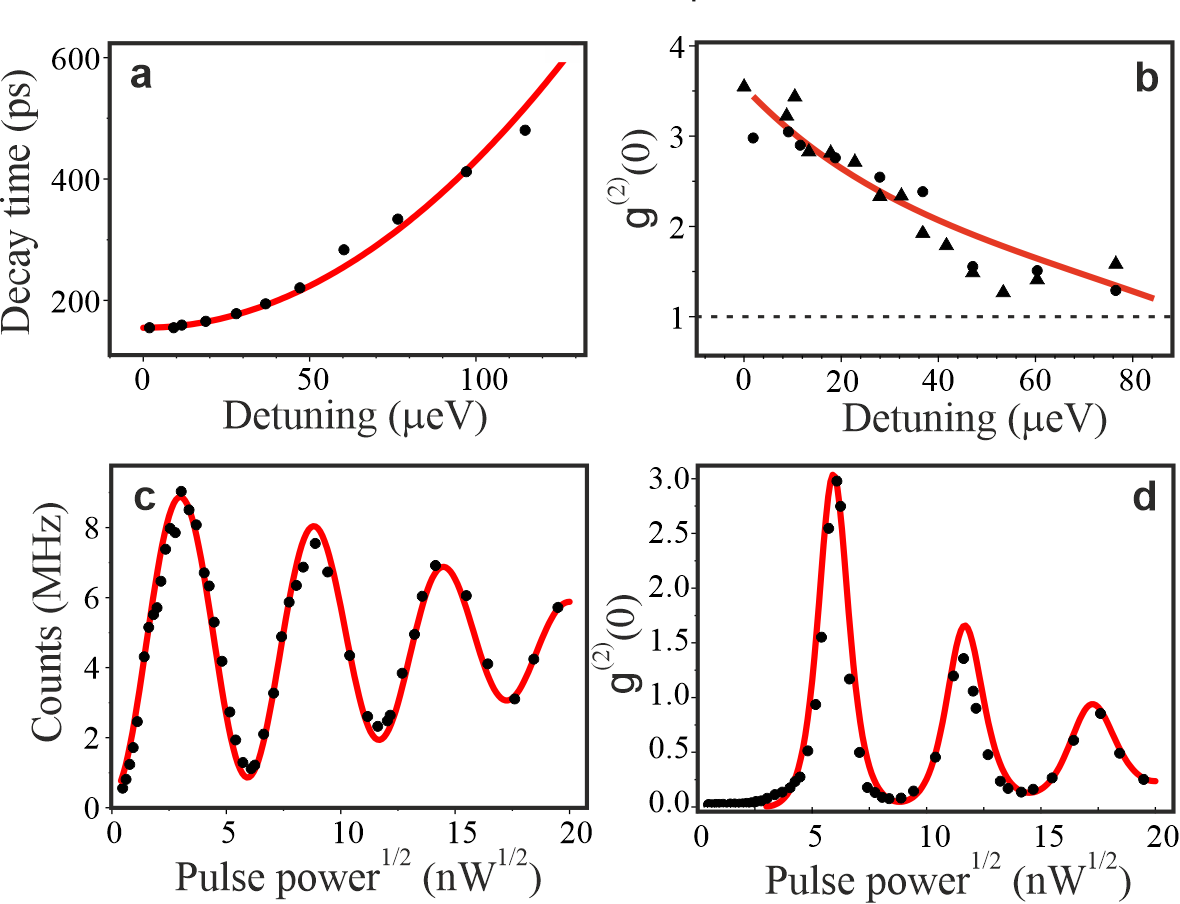}
\centering 
\caption{\label{figg2} 
Comparison of experimental data (dots) and calculated dependencies (red curves).
(a) Trion lifetime and (b) $g^{(2)}(0)$ values as a function of the detuning between the trion emission line and the optical mode. Circles and triangles in (b) represent two series of measurements.(c) Rabi oscillations in the trion emission and (d) $g^{(2)}(0)$ as a function of the pulse power. 
Simulation details are provided in the Supplementary.
}
\end{figure} 

We have developed a simple "box" model to better understand the generation of multiphoton states when resonant scattering dominates. When $\Gamma \tau \lesssim 1$, we assume that photons do not have time to leave the cavity and consider fully coherent dynamics on the scale of the pump pulse duration $\tau$. After the pulse, photons leave the cavity in time $1/\Gamma$. Then, the excited trion, if present, radiatively recombines during its lifetime. Thus, we neglect the re-excitation process (\ref{p2reEx}) and retain only resonant scattering (\ref{p2scat}) with conventional trion emission. Respectively, only the coherent term $\hat{H}_0$ in the Lindblad equation (\ref{linbEq}) is taken into account. As previously, we consider the evolution of a hole with spin up. The driving field is treated as a coherent state comprising several tens of photons ($\sim10^2$). 
For the experimental configuration as in Fig.~2a, the H-polarized pump pulse reads as:
\begin{gather}
\label{psiIn}
    \left| \Psi_{\text{in}} \right\rangle = \e^{-|\alpha|^2/2}\e^{\alpha \hat{b}^\dag_H} \hat{a}^\dag_+|0\rangle = \e^{-|\alpha|^2/2} \sum_n \frac{\alpha^n}{\sqrt{n!}} \hat{a}^\dag_+ |n_H\rangle, 
\end{gather}
where $|\alpha|^2 = N$ is the mean photon number in the pulse.


The interaction between the pump pulse and the trion is described by the evolution operator
$\hat{S}_0 = \exp \left( -i \tau \hat{H}_0 \right)$. For simplicity, we assume  that $g$ and $\alpha$ are real. The linearly polarized light $\e_H= (\sigma_+ + \sigma_-)/\sqrt{2}$. The interaction of the trion state and the $\sigma_-$ component according the selection rules leads to Rabi oscillations with frequency $\Omega_R \sim g\sqrt{n}$. The distribution of the photon number in the input pulse leads to a superposition of coherent states oscillating at different frequencies. This provides a non-zero probability of the trion state excitation for a 2$\pi$ pulse.
The outgoing state is
\begin{gather} 
\nonumber
    \left| \Psi_{\text{out}} \right\rangle = \hat{S}_0 \left| \Psi_{\text{in}} \right\rangle = 
    \e^{-|\alpha|^2/2} \e^{\alpha \hat{b}^\dag_+/\sqrt{2}} \hat{S}_0 \e^{\alpha \hat{b}^\dag_-/\sqrt{2}} \hat{a}^\dag_+ |0\rangle \approx \\ \nonumber
    \approx \frac{\e^{-|\alpha|^2/2}\e^{\alpha \hat{b}^\dag_H}}{2}  
    \biggl[
    \left(\e^{i\theta/2} \e^{i  \delta \hat{b}^\dag_-} + \e^{-i\theta/2} \e^{-i  \delta \hat{b}^\dag_-} \right) \hat{a}^\dag_+ \\ 
    - \left(\e^{i\theta/2} \e^{i  \delta \hat{b}^\dag_-} - \e^{-i\theta/2} \e^{-i \delta \hat{b}^\dag_-} \right)
    \hat{d}^\dag_+ 
    \biggl]
    |0\rangle, 
    \label{PsiOut}
\end{gather}
where $\theta = g\tau\sqrt{N/2}$ represents the pulse area, $\delta = g \tau/ 2$. Here we linearize $\sqrt{n}$ around $N$ for large mean photon numbers ($N \gg 1$). This expression shows that the wave function of the system has additions to the vacuum and trion states. Such additions are symmetric and antisymmetric Schr\"odinger cat states with phase $\delta$, the appearance of which was proposed in \cite{Loredo2019}. 
Equation (\ref{PsiOut}) highlights the difference between excitation by $\pi$ and 2$\pi$ pulses. The antisymmetric and symmetric terms are shifted in phase approximately as $\sin$ with respect to $\cos$. For a $\pi$ pulse, the antisymmetric term tends to zero, while the symmetric term tends to unity, ensuring high efficiency of single-photon emission. For the 2$\pi$ pulse, the situation changes in favor of the antisymmetric term, and the scattered signal increases.


After the pulse excitation, the trion emits a $\sigma_-$ photon, so we can replace the trion creation operator with a photon  $\hat{d}^\dag_+ \rightarrow -i \hat{b}^\dag_{t-}$  (see the Supplementary for details). 
The last term in (\ref{PsiOut}) contains the contribution of resonant scattering when the trion is in an excited state and the photon is in a mode, which is described by the bottom line in expression (\ref{p2scat}) within the full model.
The detection scheme uses a cross-polarized (V) projection of the output state:
\begin{gather} \nonumber
    \left| \Psi_{\text{out}}^{\text{V}} \right\rangle \approx \cos\left(\frac{\theta}{2} \right) \left| 0 \right\rangle +  \frac{1}{\sqrt{2}} \sin\left(\frac{\theta}{2} \right) \left( \left| 1_t \right\rangle + i\delta\, \left| 1_V \right\rangle \right) \\ - \frac{i\delta}{2} \cos\left(\frac{\theta}{2} \right) \left| 2_{t,V} \right\rangle,
    \label{PsiOutSimlpe}
\end{gather}
where we retain only first-order terms in $\delta \sim g\tau \ll 1$. The state $\left| 1_t \right\rangle$ is the photon from trion recombination, $\left| 1_V \right\rangle$ is the laser photon resonantly scattered by the trion, and $\left| 2_{t,V} \right\rangle$ is a state with two photons, one of which is the laser photon resonantly scattered by the trion and the other is the photon resulting from trion recombination. Note that these two-photon states oscillate coherently with the pulse area, participating in the Rabi oscillations in the second-order correlation function.
For the output state (\ref{PsiOutSimlpe}), the $g^{(2)}(0)$ is given by:
\begin{gather}
    g^{(2)}(0) = \frac{ \left\langle \Psi_{\text{out}}^{\text{V}} \right| \hat{n}^2 - \hat{n}\left| \Psi_{\text{out}}^{\text{V}} \right\rangle}{\left\langle \Psi_{\text{out}}^{\text{V}} \right| \hat{n} \left| \Psi_{\text{out}}^{\text{V}} \right\rangle^2} = \frac{2|\delta|^2 \cos^2\left(\frac{\theta}{2} \right) }{ \left(\sin^2\left(\frac{\theta}{2} \right) + |\delta|^2
    \right)^2}.
    \label{g2Simple}
\end{gather} 
Equation (\ref{g2Simple}) shows the oscillatory dependence of $g^{(2)}(0)$ on the pump pulse area. Notably, $g^{(2)}(0)$ vanishes for odd $\pi$ pulse areas ($\pi$, $3\pi$, ...) and peaks for even $\pi$ areas ($2\pi$, $4\pi$, ...). This behavior demonstrates that the multiphoton states are the result of fully coherent dynamics. 
Note that this equation, derived under the assumption $N \gg 1$, is invalid for pulse areas close to zero, where coherence is destroyed. For a fixed pump power, the correlation function exhibits damped oscillations with increasing pulse duration $\tau$. For a fixed $\tau$, it should exhibit undamped oscillations with increasing pump power.

The experimentally observed decrease in $g^{(2)}(0)$ with increasing pump power at a fixed $\tau$ indicates the influence of  phonons \cite{Rabi_Dump_Exp_2010,Nazir_2016}. 
In the Supplementary, we present the solution of the master Lindblad equation for the "box" model, where phonons determine the rate of phonon-induced decay of $\gamma$ and a small correction $\varepsilon$ to the Rabi frequency \cite{Rabi_Shift_Exp_2010}. We consider the high-temperature regime ($T \gg \Omega_R$), in which $\gamma \sim \Omega_R^2$ leads to the decay of both $g^{(2)}(0)$ and Rabi oscillations in the slow trion PL $P_{\text{trion}} = [1 - \e^{-2\gamma \tau} \cos (\tilde{\theta}) ]/2$, where $\tilde{\theta} = \theta - \varepsilon \tau$. Examples of modeling the experimental data from Fig.~2d using the improved model are shown in Figs.~3c,d.

For a neutral exciton with degenerate levels, excitation with successful filtering in a cross-polarization scheme is impossible. When the  anisotropy of the QD structure lifts this degeneracy, leading to a nonzero fine-structure splitting $\Delta_{FSS}$ between two orthogonally polarized levels, the exciton can generate multiphoton states to the extent of $\tau\Delta_{FSS} $ and similarly control the nonlinear polarization rotation.
In the time domain, the fine-structure splitting manifests itself as a periodic modulation of the PL decay curves. As a result, the first maximum, corresponding to the most efficient excitation, does not coincide in time with the peak of the laser pulse  (see Fig.~S3 in the Supplementary). In this case, the resonant scattering becomes less efficient, and the probability of generating a photon pair is significantly reduced. The measured value of $g^{(2)}(0)$ determined by eq. (3) becomes sub-Poissonian, as shown in Fig.~S4 in the Supplementary.


In conclusion, we investigated the nature of the multiphoton bunch formed upon resonant excitation of a charged  QD placed in a cavity. This was done by analyzing the Rabi oscillations in the PL components characterized by different fast and slow decay  and comparing them with the change in the function $g^{(2)}(0)$,  measured at the same power values up to $6\pi$. In addition, the dependence of $g^{(2)}(0)$ was measured with changing the pulse duration and with detuning the trion transition energy with respect to the resonator mode.  These experiments showed that the multiphoton bunch is a coherent superposition of resonantly scattered and emitted photons. They also demonstrated that an increase in the density of photon states upon placing a QD in an optical cavity affects the quantum statistics of photons in such bunches.
We developed a theory describing the generation of the multiphoton bunches, which takes into account the contribution of re-excitation and resonance scattering mechanisms, the presence of a trion in the QD, and coupling with optical modes.  For short pump pulses, we developed a simple "box" model describing the resonant scattering of the photon states that comprise the laser pulse.  This model proved to be applicable for modeling the obtained experimental data. In general, our approach bridges the Jaynes-Cummings dynamics with the physics of Mollow resonance scattering. This approach describes a path to a comprehensive study of multiphoton states, which are high-order elements of the Fock space and hold promise for quantum technologies. \cite{Mollow1969}


\bibliography{main}

\end{document}


\setcounter{figure}{0}            
\renewcommand{\thefigure}{S\arabic{figure}}
\title{Supplemental Material \\ 
Coherent superposition of emitted and resonantly scattered photons from a two-level system in a cavity driven by an even-$\pi$ pulse
}

\author{I.V.~Krainov}
\email{igor.kraynov@mail.ru} 
\affiliation{Ioffe Institute, 194021 St.~Petersburg, Russia} 
\author{A.I.~Galimov}
\affiliation{Ioffe Institute, 194021 St.~Petersburg, Russia} 
\author{Yu.M.~Serov}
\affiliation{Ioffe Institute, 194021 St.~Petersburg, Russia} 
\author{A.A.~Toropov}
\affiliation{Ioffe Institute, 194021 St.~Petersburg, Russia} 
\author{T.V.~Shubina}
\affiliation{Ioffe Institute, 194021 St.~Petersburg, Russia}

\maketitle

\section{Experimental details}

The sample under study is a high-quality single-photon emitter with a cylindrical micropillar of 3 $\mu$m in diameter, made of a heterostructure with distributed Bragg reflectors (DBR) grown by molecular beam epitaxy. It has 30 (18) bottom (top) pairs of $\lambda$/4 GaAs/AlGaAs layers and a positively charged InAs/GaAs QD in the $\lambda$-cavity between them. Technology details can be found in \cite{Galimov2025}. During the experiments, the sample was placed in a helium flow cryostat with a minimum temperature of 8 K. 
According to the reflectance spectra measured from the micropillar tip (Fig. S1 (a)), the cavity mode width is $\sim$118 $\mu$eV. The polarization splitting of the fundamental mode is negligible ($<$10 $\mu$eV) compared to the spectral mode width; therefore, it cannot lead to polarization rotation of the scattered laser photons. The Purcell factor for the microcavity is $\sim$5.4. The QD PL line with a maximum at 917.64 nm spectrally coincides with the microcavity mode at a sample temperature of 9.4 K (Fig. S1). The PL spectra were measured using a triple-grating spectrometer with a spectral resolution of about 10 $\mu$eV. The observed linewidth of the QD PL in Fig.~S1b is limited by this resolution. 

\begin{figure*} 
\includegraphics[width=0.5\linewidth]{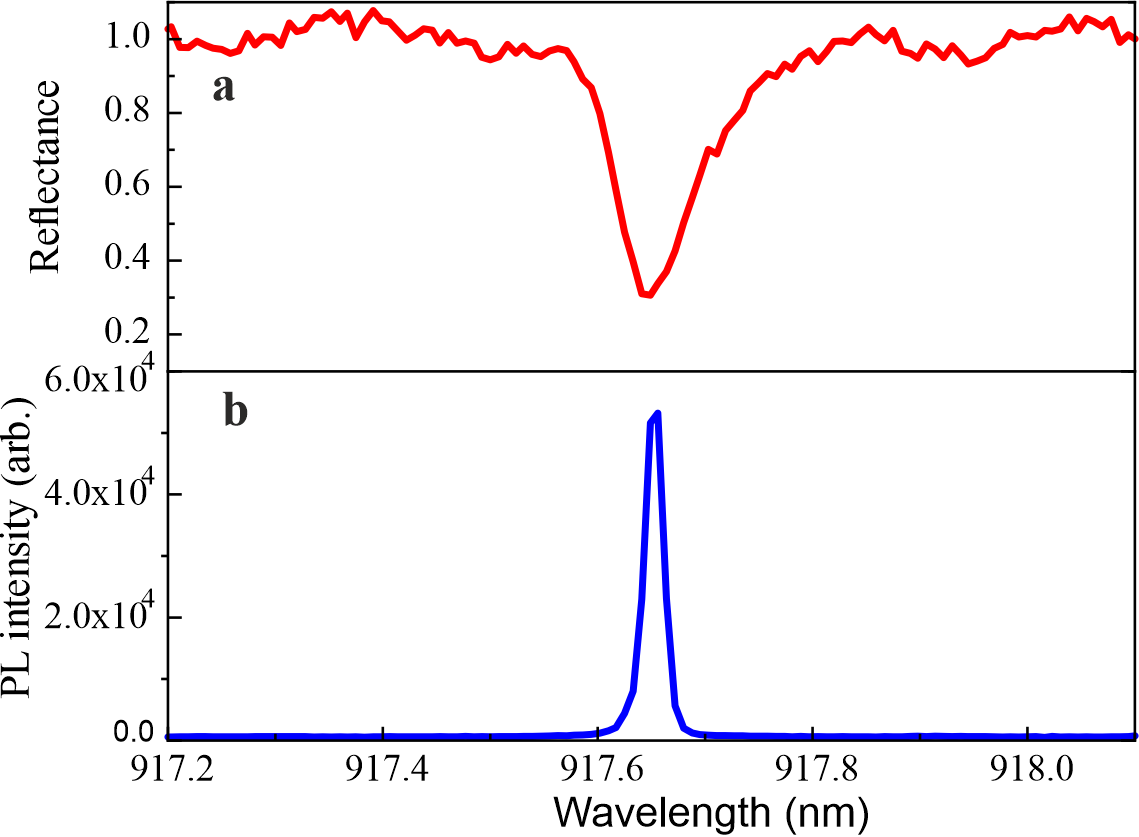}
\centering
\caption{
(a) Reflectance spectrum of a micropillar at 9.4 K. (b) Photoluminescence spectrum of a QD embedded into the micropillar.}
\label{fig1S}
\end{figure*}

For resonant excitation, femtosecond pulses from a parametric laser oscillator were transformed using a 4f pulse shaper \cite{Monmayrant2010} to fine-tune the wavelength and pulse duration. The chosen pulse duration of 16 ps ensures optimal laser filtering efficiency and single-photon purity. The spectral width of the laser pulse is $\sim$80 $\mu$eV that is narrower than the cavity mode width and allows the laser pulse to propagate without distortion.  
An estimate of the number of photons in a 16-ps $\pi$-pulse with a power of 9 nW outside the cryostat  and a repetition rate of 78.3 MHz shows that the number of photons incident on the micropillar tip is about 500. However, if we take into account the reflectivity spectrum and the small spectral overlap of the laser pulse width ($\sim$80 $\mu$eV) with the linewidth of the QD ($<$10 $\mu$eV), the effective number of photons should decrease several times. To refine this estimate, we use another method, "from the inside". For excitation by the $\pi$-area pulse: $\pi = g\tau_p\sqrt{N}/\hbar$, where $g$ is the coupling strength, $\tau_p$ is the pulse duration. For our InAs QD in the micropillar cavity $g\approx$ 20$\pm$2 $\mu$eV, which is close to that observed in \cite{Giesz_2016}. This corresponds to approximately 50 photon states in a $\pi$-pulse. A realistic variation of $g$ within 10-20 $\mu$eV gives a possible range of 50-100 photons.

Measurements of the second-order correlation function $g^{(2)}(\tau)$ were performed using a Hanbury-Brown-Twiss interferometer equipped with superconducting nanowire single-photon detectors.
Examples of the original $g^{(2)}(\tau)$ dependencies are shown in Fig. S2. The area under the curve represents the absolute value of $g^{(2)}$. Normalization was performed using the average of peak areas over sufficiently long time intervals to ensure the absence of correlation and $g^{(2)}(\infty)=1$.  This procedure was used to obtain the histogram shown in Fig. 2b in the main text.

\begin{figure*}
\includegraphics[width=0.9\linewidth]{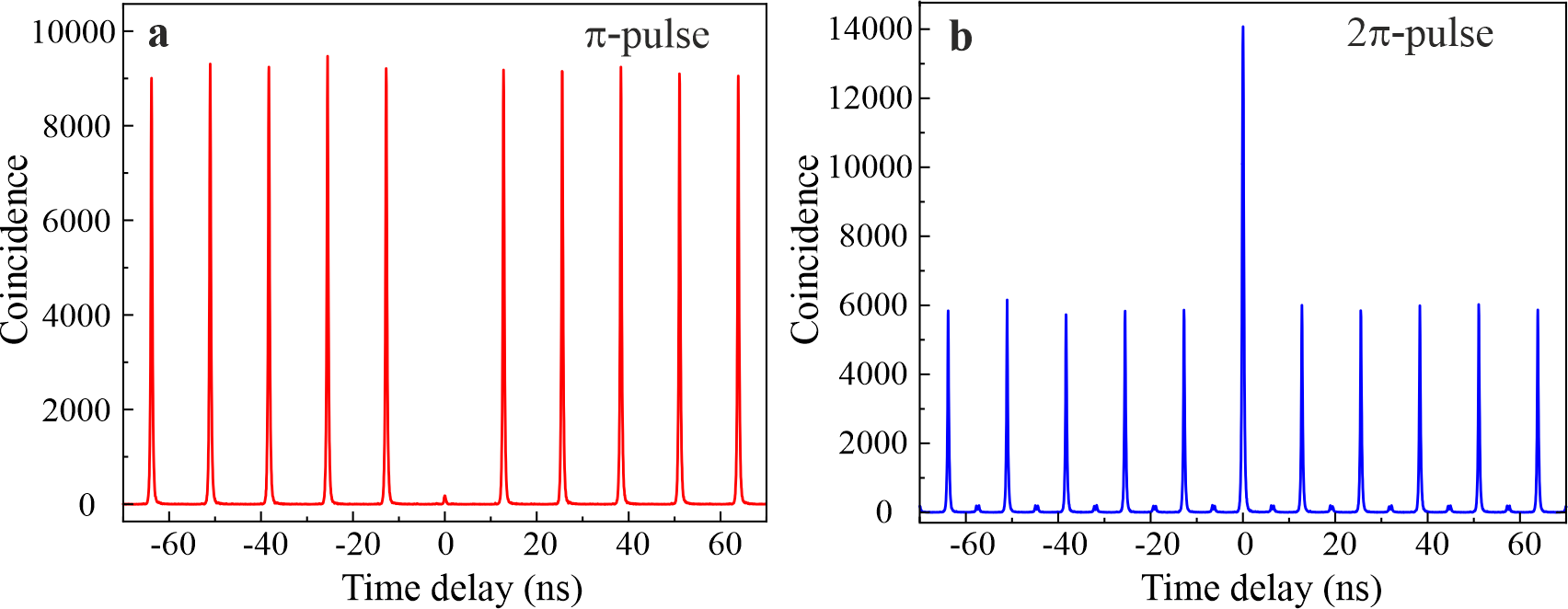}
\centering
\caption{
Examples of the original experimental data of $g^{(2)}(\tau)$ of QD emission excited by a $\pi$-pulse (a) and a 2$\pi$-pulse (b).}
\label{fig2S}
\end{figure*}

To obtain correct Rabi oscillations for theoretical modeling, we eliminated parasitic factors. The InAs/GaAs QDs suffer from blinking even under resonant excitation \cite{Nguyen2013}. This leads to a seeming increase in the peak values of the Rabi oscillations, as previously observed in \cite{Fischer2017,Loredo2019}. To reconstruct the true Rabi oscillation profile, we quantified the QD bright state occupancy probability for each power using long-term autocorrelation function measurements \cite{Hilaire2020} and normalized the raw data by this parameter. These corrected graphs are shown in Fig. 2d and Fig. 2e in the main text. The time-resolved PL decay curve (Fig. 2c) is measured using the time-correlated single-photon counting technique. The PL and laser emission spectra, shown in the inset to Fig. 2c, were measured using a triple-grating spectrometer (see above). 

To obtain the dependencies of the intensities of scattered laser radiation and QD PL on the pulse power (Fig. 2d), we measured the decay curves for each power value. The corresponding count rates are extracted from the decay curves by fitting with a function $f(t)$ that includes a fast component associated with the laser and a slowly-decaying component of the trion emission: 
$$f(t)=A_{laser} \e^{-t^2/2\tau^2}+A_{PL} \e^{-t/\tau_{tr}},$$
where $\tau_{tr}$ is the trion PL decay time (155 ps), $\tau$ is the laser pulse duration (16 ps),  $A_{PL}$ and $A_{laser}$ are the amplitudes of the PL signal and the scattered laser signal, respectively. 
The corresponding timing uncertainty of the detection setup ($\delta_{ds}$) is 39 ps. This value accounts for all setup uncertainties, including those associated with the detectors, time tagger, and electrical cables. To correctly describe the experimental data, the function $f(t)$ was convolved with the instrument response function $\exp(-t^2/2\delta_{ds}^2)$. The size of the time bins was chosen to accurately reflect the results of time-resolved PL measurements; for this experiment, it was set at 10 ps.

In the QD-cavity detuning experiments (Fig. 3), the temperature was increased by only about 10 K to $\sim$20 K to prevent significant disruption of the photon emission process. The pulse area and duration were kept constant and the laser photon energy followed the QD line energy. To measure the trion lifetime unaffected by the Purcell effect, we increased the temperature to 37 K to achieve complete decoupling between the QD and the cavity.

\section{Generation of multiphoton states by a neutral exciton}

As noted in the main text, for a nonzero fine-structure splitting $\Delta_{FSS}$ between two orthogonally polarized levels, a neutral exciton can generate multiphoton states and rotate the polarization to the extent of an amount $\tau\Delta_{FSS} $ ($\tau$ is the pulse duration). The small value of FSS hampers these effects.  An additional suppressing factor is the mismatch in the time domain of the laser pulse and the first maximum of the "beats" between two orthogonally polarized levels. The smaller the value of FSS, the greater this temporal mismatch (Fig. S3). As a result, both resonant scattering of laser photons and re-excitation become ineffective, and $g^{(2)}(0) \sim 0.1 $ is sub-Poissonian (Fig. S4).
\begin{figure*}
\includegraphics[width=0.45\linewidth]{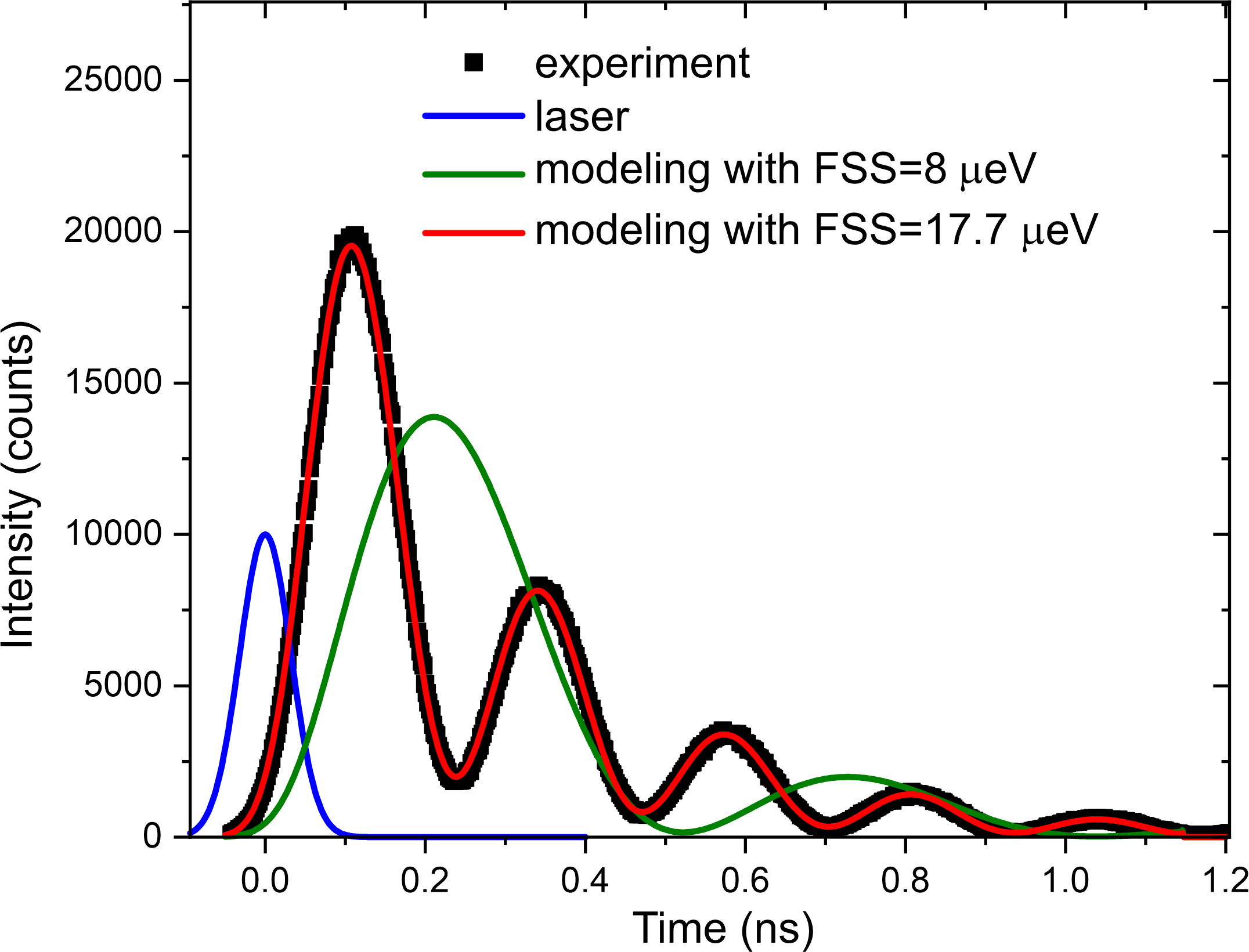}
\centering
\caption{
Experimental and simulated beatings between two orthogonally polarized levels of a neutral exciton are shown along with the laser pulse. The FSS values used for simulations are specified in the plot. }
\label{ExcitonFSS}
\end{figure*}

\begin{figure*}
\includegraphics[width=0.9\linewidth]{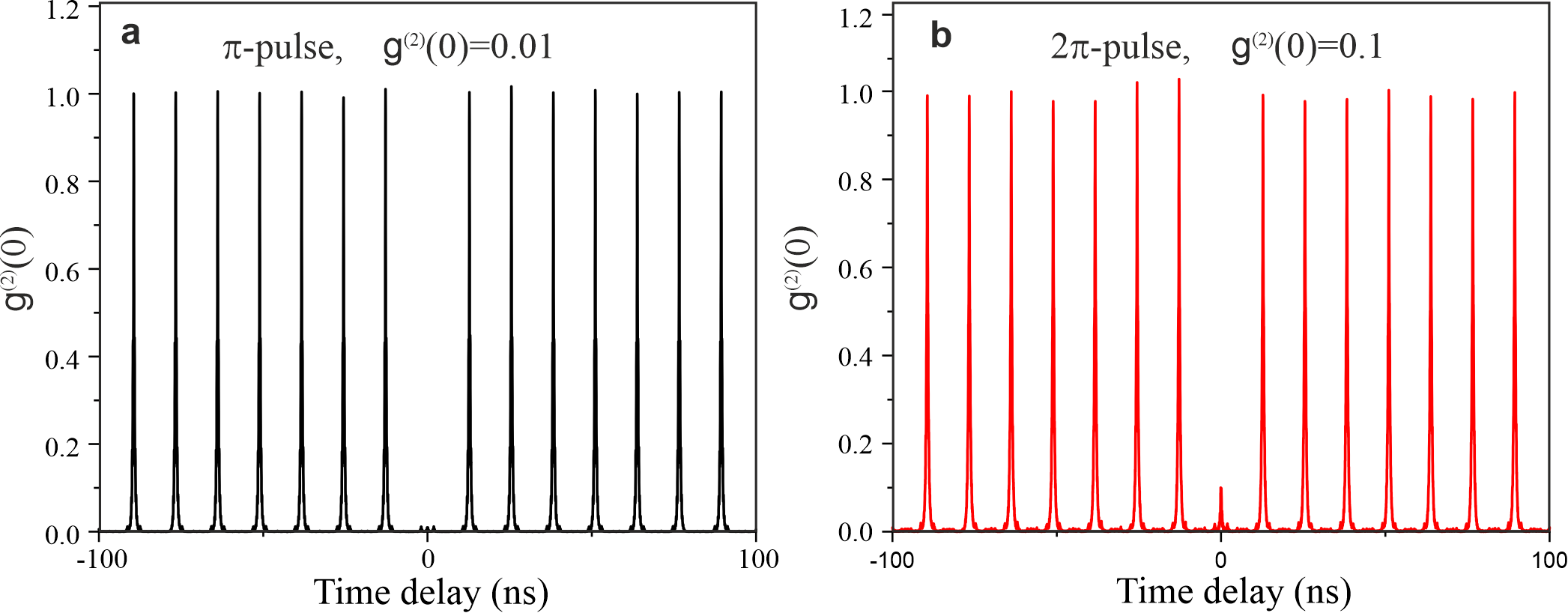}
\centering
\caption{
$g^{(2)}(\tau)$ for QD emission excited by a $\pi$-pulse (a) and a 2$\pi$-pulse (b).}
\label{ExcitonG2}
\end{figure*}

\section{Comparison of resonant scattering and re-excitation terms}
To compare the contributions of the resonant scattering and re-excitation terms to two-photon generation, we can evaluate them using the non-Hermitian Hamiltonian \cite{Carmichael} describing the dynamical Lindblad equation (Eq.~(2) in the main text). For simplicity, in this section we remove phonons and all other imperfections from the Lindbladian. In the H polarized pumping pulse, we replace the photon operators with the classical field $\Omega_R(t)$. As in the main text, we assume that the resident hole in the QD is initially in a spin up state, i.e. $|\Psi(0)\rangle = \hat{a}^\dag_+ |0\rangle $ and write only the spin up part of the Hamiltonian; for the opposite spin, everything is similar. The non-Hermitian Hamiltonian under consideration is: 
\begin{gather}
\label{nonHermHam}
    \hat{H} = \hat{\mathcal{H}}_0 + \hat{\mathcal{V}}, \quad 
    \hat{\mathcal{H}}_0 = \Omega_R(t)\, (\hat{a}^\dag_+ \hat{d}_+ +  \hat{a}_+ \hat{d}^\dag_+ )  
    - i \frac{\Gamma}{2} \hat{b}^\dag_V \hat{b}_V, \quad
    \hat{\mathcal{V}} = 
    \frac{g}{\sqrt{2}} ( -i\hat{b}^\dag_V \hat{a}^\dag_+ \hat{d}_+ + i \hat{b}_V \hat{a}_+ \hat{d}^\dag_+ ).
\end{gather}
It is assumed that the relationship between the parameters is as follows: $g\ll \Gamma$, $g\ll \Omega_R$ ($\hbar=1$). To detect two emitted photons, it is necessary to sum up all the possibilities of their detection at different moments in time:
\begin{gather}
\label{P2def}
    P_2 = \Gamma^2 \int_0^\infty dt_2 \int_0^\infty dt_1 \text{Tr} \left[  \hat{b}_V \hat{S}(t_2, t_1) \hat{b}_V \hat{S}(t_1, 0) |\Psi(0)\rangle 
    \langle \Psi(0) | \hat{S}(0, t_1) \hat{b}^\dag_V \hat{S}(t_1, t_2) \hat{b}^\dag_V 
    \right].
\end{gather}
For simplicity, we assume that the pulse has a square shape: $\Omega_R(t) = \Omega$ for $t \in (0,\tau)$ and $\Omega_R(t) = 0$ otherwise. 
In this case, we can split the matrix S at large times into two parts — at the moment of the pulse and after the pulse: $\hat{S}(\infty, 0) = \hat{S}(\infty, \tau) \hat{S}(\tau, 0)$. For short pulses, we can obtain the matrix S in perturbation theory on $g$, so that the wave function has the form:
\begin{gather}
    |\Psi(t)\rangle = \text{T} \exp \left( -i \int_\tau^t d\tau' \hat{H}(\tau') \right)
    \e^{-i \hat{\mathcal{H}}_0 \tau} \left( 1 - i \int_0^\tau d\tau' \e^{i \hat{\mathcal{H}}_0 \tau'} \hat{\mathcal{V}} \e^{-i \hat{\mathcal{H}}_0 \tau'} \right) |\Psi(0)\rangle.
\end{gather} 
The re-excitation term arises when the system at the pulse emits a photon and drops into the initial state (ground state), so that the state $\hat{b}_V \hat{S}(t_1, 0) |\Psi(0)\rangle$ is projected into the state $\hat{a}^\dag_+ |0\rangle$, see process (6) in the main text: 
\begin{gather}
\label{reexProj}
\langle0| \hat{a}_+ \hat{b}_V \left( - i \int_0^{t_1} d\tau' \e^{-i \hat{\mathcal{H}}_0 (t_1-\tau')} \hat{\mathcal{V}} \e^{-i \hat{\mathcal{H}}_0 \tau'} \right) \hat{a}^\dag_+ |0\rangle = \frac{g (-4\Omega + 4 \Omega \e^{-\Gamma t/2} \cos(2\Omega t) + \Gamma \e^{-\Gamma t/2} \sin(2\Omega t))}{\sqrt{2}(\Gamma^2+16\Omega^2)}.
\end{gather}
After this, we need to re-excite the system to produce two photons. 
As we obtain a result on the minimal $g$ order, we will neglect it for the second part of the pulse:
\begin{gather}
\e^{-i \hat{\mathcal{H}}_0 (\tau-t_1)} \hat{a}^\dag_+ |0\rangle\langle0| \hat{a}_+ \hat{b}_V \left( - i \int_0^{t_1} d\tau' \e^{-i \hat{\mathcal{H}}_0 (t_1-\tau')} \hat{\mathcal{V}} \e^{-i \hat{\mathcal{H}}_0 \tau'} \right) \hat{a}^\dag_+ |0\rangle.
\end{gather}
After the pulse, the Hamiltonian describes that all cavity photons leave the resonator with rate $\Gamma$ and the trion state emits photons with rate $g^2/\Gamma$. Thus, the second integral over $t_2$ with multiplication by the second V-photon $\hat{b}_V$ in (\ref{P2def}) can be simplified to a projection onto it, and we obtain a term with two photons arising in the re-excitation process with probability:
\begin{gather}
P_2^{re\, ex} = \Gamma \int_0^\tau dt_1 
\left| \langle0| \hat{d}_+\e^{-i \hat{\mathcal{H}}_0 (\tau-t_1)} \hat{a}^\dag_+ |0\rangle \langle0| \hat{a}_+ \hat{b}_V \left( - i \int_0^{t_1} d\tau' \e^{-i \hat{\mathcal{H}}_0 (t_1-\tau')} \hat{\mathcal{V}} \e^{-i \hat{\mathcal{H}}_0 \tau'} \right) \hat{a}^\dag_+ |0\rangle \right|^2.
\end{gather}
The first matrix element in this equation must be calculated to zeroth order in $g$, and it is equal to $|\langle0| \hat{d}_+\e^{-i \hat{\mathcal{H}}_0 (\tau-t_1)} \hat{a}^\dag_+ |0\rangle|^2 = \sin^2(\Omega (\tau-t_1))$ - the probability of re-excitation.
The other branch of the dynamics is described by the resonant scattering term, see the process (7) in the main text. The main difference with the re-excitation term is that the system has a resonantly scattered photon and an excited trion. If we evaluate this possibility at pulse times, we need to project the wave function onto the trion state instead of the ground state in (\ref{reexProj}):
\begin{gather}
\langle0| \hat{d}_+ \hat{b}_V \left( - i \int_0^{t_1} d\tau' \e^{-i \hat{\mathcal{H}}_0 (t_1-\tau')} \hat{\mathcal{V}} \e^{-i \hat{\mathcal{H}}_0 \tau'} \right) \hat{a}^\dag_+ |0\rangle = 
\frac{i g \e^{-\Gamma t/2} (\Gamma^2 + 16 \Omega^2(1-\e^{\Gamma t/2}) - \Gamma^2 \cos(2\Omega t) + 4 \Gamma \Omega  \sin(2\Omega t))}{\sqrt{2}\Gamma (\Gamma^2+16\Omega^2)}.
\end{gather}
After the resonant scattering process, the trion must remain in an excited state in the remaining part of the pulse to emit a second photon, and we obtain the two-photon resonant scattering term:
\begin{gather} \nonumber
P_2^{scat} = \Gamma \int_0^\tau dt_1 
\left| \langle0| \hat{d}_+\e^{-i \hat{\mathcal{H}}_0 (\tau-t_1)} \hat{d}^\dag_+ |0\rangle \langle0| \hat{d}_+ \hat{b}_V \left( - i \int_0^{t_1} d\tau' \e^{-i \hat{\mathcal{H}}_0 (t_1-\tau')} \hat{\mathcal{V}} \e^{-i \hat{\mathcal{H}}_0 \tau'} \right) \hat{a}^\dag_+ |0\rangle \right|^2 + \\
+ \left| \langle0| \hat{d}_+ \hat{b}_V \left( - i \int_0^{\tau} d\tau' \e^{-i \hat{\mathcal{H}}_0 (\tau-\tau')} \hat{\mathcal{V}} \e^{-i \hat{\mathcal{H}}_0 \tau'} \right) \hat{a}^\dag_+ |0\rangle \right|^2,
\end{gather}
where the first line ($P_2^{scat\, 1}$) describes the probability that a resonant scattering photon is emitted during the pulse length, and the second line ($P_2^{scat\, 2}$) indicates that the resonant scattering photon and the trion emission are detected after the pulse. The first matrix element in the first line of this equation must be calculated to the zeroth order in $g$, and it is equal to $|\langle0| \hat{d}_+\e^{-i \hat{\mathcal{H}}_0 (\tau-t_1)} \hat{d}^\dag_+ |0\rangle|^2 = \cos^2(\Omega (\tau-t_1))$ - the probability of a trion remain in an excited state. Then, for short pulses, one must set $\Omega = \pi \Gamma$ and $\tau = 1/\Gamma$ to obtain the ratio given in the main text. It is also interesting to compare the terms of resonant scattering in this case: $P_2^{scat\, 2}/P_2^{scat\, 1} \approx 4.3$.

\section{Trion dynamics with phonons in the "box model”}

We will analyze the dynamic damping of a trion by phonons, as described in the main text, assuming a "box model". This model assumes the presence of phonons only in the damping term of the Lindblad equation. Phonons also introduce a small correction to the oscillations frequency, so in this section we derive the Lindbladian. 
The full Hamiltonian describing the trion dynamics under light pumping in the presence of acoustic phonons consists of three components:
\begin{gather}
    \hat{H} = \hat{H}_0 + \hat{H}_p + \hat{H}_i, \\
    \hat{H}_p = \sum_k \omega_k \hat{c}^\dagger_k \hat{c}_k, \quad 
    \hat{H}_i = (\hat{d}_+^\dagger \hat{d}_+ + \hat{d}_-^\dagger \hat{d}_-)\sum_k t_k (\hat{c}^\dagger_k + \hat{c}_k),
\end{gather}
where $t_k$ is the trion-phonon coupling constant, $\hat{c}^\dagger_k$ creates phonons in the mode $k$ with dispersion $\omega_k = c_p k$.
Following the established procedure \cite{Nazir_2016}, we treat phonon-induced dephasing using second-order perturbation theory. For simplicity, we assume $\hbar = 1$ in the equations. Moving on to the picture of the interaction with respect to $\hat{H}_0$ and $\hat{H}_p$ and tracing over the phonon bath, we obtain the density matrix equation:
\begin{align}
    \frac{\partial \hat{\rho}_t}{\partial t} &= -i \left[ \hat{H}_0 ; \hat{\rho}_t \right] - \int^t_{-\infty} d\tau 
    \text{Tr}_p \left( 
    \left[ \hat{H}_i (t, 0);
    \left[
    \hat{H}_i (\tau, \tau - t) ; 
    \hat{\rho}_t \hat{\rho}_p 
    \right] \right]
    \right), \\
    &\hat{H}_i (\tau, t) = \e^{i ( \hat{H}_0 t + \hat{H}_p \tau)} \hat{H}_i \e^{-i ( \hat{H}_0 t + \hat{H}_p \tau)}.
    \nonumber
\end{align}
When the phonon-driven decay rate remains small compared to the Rabi frequency ($\gamma \ll \Omega_R$), we can replace the photon operators in the decay terms with their average values, effectively using the classical Rabi frequency $\Omega_R = g\sqrt{N}$. As in main text, we assume that the resident hole located in the QD has a spin up state, i.e. $\hat{\rho}_0 = \hat{a}^\dag_+ |0\rangle \langle 0 | \hat{a}_+$ (the same can be obtained for the opposite spin). The density matrix can be represented in the non-trionic/trionic subspace as a $2 \times 2$ matrix in this case, where each element contains states with different number of photons. The master Lindblad equation then takes the form:
\begin{gather}
\label{eqDensMatr}
\frac{\partial \hat{\rho}_t}{\partial t} = -i \left[ \hat{H}_0, \hat{\rho}_t \right] - \gamma \left[ \hat{\sigma}_z, \left[\hat{\sigma}_z, \hat{\rho}_t \right] \right] + \varepsilon \left[ \hat{\sigma}_z, \left[\hat{\sigma}_y, \hat{\rho}_t \right] \right], \\
\gamma = \int_0^\infty \frac{d\tau}{2} \cos(\Omega_R \tau) \int d\omega \, J_p(\omega) n_\omega \cos(\omega \tau), \quad
\varepsilon = \int_0^\infty \frac{d\tau}{2} \sin(\Omega_R \tau) \int d\omega \, J_p(\omega) n_\omega \cos(\omega \tau),
\end{gather}
where, $J_p(\omega) = \sum_k t_k^2 \delta(\omega - \omega_k)$ represents the trion-phonon coupling strength \cite{Nazir_2016}, $n_\omega$ is the phonon distribution function.
In the absence of dephasing, the solution reduces to the coherent case $\hat{\rho} = \hat{S}_0 \hat{\rho}_0 \hat{S}_0^\dagger$ discussed in the main text, with $\hat{S}_0 = \exp (-i\tau \hat{H}_0)$. We assume a high-temperature regime ($T \gg \Omega_R$), then the factor $\gamma \sim T \Omega_R^2 $ describes the phonon-induced damping of the Rabi oscillations. The main problem arises from the mixing of states with different photon numbers. In the case of weak dephasing ($\gamma, \varepsilon \ll \Omega_R$), we account for it by transforming to the interaction picture $\hat{\rho} = e^{-i \hat{H}_0 t} \hat{\chi} e^{i \hat{H}_0 t}$, which gives:
\begin{equation}
\frac{\partial \hat{\chi}}{\partial t} = 
- e^{i \hat{H}_0 t}  \left( \gamma \left[ \hat{\sigma}_z, \left[\hat{\sigma}_z, e^{-i \hat{H}_0 t} \hat{\chi} e^{i \hat{H}_0 t} \right] \right] - \varepsilon \left[ \hat{\sigma}_z, \left[\hat{\sigma}_y, e^{-i \hat{H}_0 t} \hat{\chi} e^{i \hat{H}_0 t} \right] \right] \right) e^{-i \hat{H}_0 t},
\end{equation}
Averaging over the period of Rabi oscillations $\theta$ with small damping gives:
\begin{gather}
    \frac{\partial \hat{\chi}}{\partial t} = i \varepsilon \left[ \hat{\sigma}_x, \hat{\chi} \right] - \gamma \left( 
    \begin{matrix}
        \chi_{11} - \chi_{22} && 3 \chi_{12} + \chi_{21} \\
        3 \chi_{21} + \chi_{12} && \chi_{22} - \chi_{11}
    \end{matrix}
    \right).
\end{gather}
The term $\varepsilon$ can be eliminated by redefining $\hat{H}_0 \rightarrow \hat{H}_0 - \varepsilon \hat{\sigma}_x$, which corresponds to a small correction of the Rabi frequency. 
Defining the incident pump coherent light as $\left|\alpha_H\right\rangle = \exp(-|\alpha|^2/2 + \alpha \hat{b}^\dag_H) \left|0\right\rangle$, the initial condition $\chi_{11}(0) = \left|\alpha_H\right\rangle \left\langle\alpha_H\right|$, and all other terms equal to zero, we obtain the solution of the density matrix after the pump pulse:
\begin{gather} \label{rhoFinal}
    \hat{\rho}_{\text{out}} = \text{Tr}_2\,  \hat{S}_1 \frac{1}{2} \left( \hat{I} + \hat{\sigma}_z \e^{-2\gamma \tau}  \right)
    \left( \begin{matrix}
    \left|\alpha_H\right\rangle \left\langle\alpha_H\right| && 0 \\
    0 && \hat{b}^\dag_{t-} \left|\alpha_H\right\rangle \left\langle\alpha_H\right| \hat{b}_{t-} 
    \end{matrix} \right)
    \hat{S}_1^\dagger, \\
    \hat{S}_1 = \exp\left[ -i \tau \left(
     \begin{matrix}
        0 && g \hat{b}_- - \varepsilon \\
        g \hat{b}^\dag_- - \varepsilon && 0
    \end{matrix}
    \right) \right],
    \nonumber
\end{gather}
where trace is made over this 2x2 matrix, $\hat{b}^\dag_{t-}$ is the trion emitted photon creation operator.
Using $\hat{\rho}_{\text{out}}$ we derive the second-order correlation function: 
\begin{gather}
\label{g2Final}
    g^{(2)}(0) = \frac{
    4 |\delta|^2 \left( 1 + \e^{-2\gamma \tau} \cos(\tilde{\theta}) \right)
    }{
    \left(
    1 - \e^{-2\gamma \tau} \cos(\tilde{\theta}) + 2 |\delta|^2 
    \right)^2 },
\end{gather} 
where $\tilde{\theta} = \theta - \varepsilon \tau$. The quadratic dependence $\gamma \sim \theta^2$ explains the observed decay of $g^{(2)}(0)$ with increasing pump power, as well as the damping of Rabi oscillations:
\begin{gather}
    P_{\text{trion}} = \frac{1}{2} [1 - \e^{-2\gamma \tau} \cos (\tilde{\theta}) ].
\end{gather}

\section{Dependence on detuning}

To verify our model, we investigate its ability to describe multiphoton states when the interaction between the trion and the cavity mode varies. To do so, we introduce an energy detuning between the cavity mode and the trion transition, defined as $\hbar \omega_0 - E_t$, keeping the pulse area fixed and resonant excitation $\hbar \omega_L = E_t$, which corresponds to the experimental conditions. For $\theta = 2\pi$, the main difference from the ideal case is that the phonon induced dephasing limits the denominator: 
\begin{gather}
    g^{(2)}(0) \approx \frac{2|\delta|^2}{\left( |\delta|^2 + \gamma  \tau \right)^2}.
\end{gather}
According to this equation, the detuning affects both important parameters. The interaction strength $\delta$ depends on the detuning due to the changes in the photon density of states. The trion decay rate $\tau_t$ depends on the photon density of states as follows: 
\begin{gather}
\frac{1}{\tau_t} \sim \frac{g^2 \Gamma}{\Gamma^2 + (\omega_0 - E_t)^2}.
\end{gather}
By measuring and verifying this dependence, we obtain a scaling of the interaction strength depending on the detuning: $\delta^2(\omega_0 - E_t) / \delta^2(0) = \tau_t(0) / \tau_t(\omega_0 - E_t)$. 
Another important factor is the change in the phonon dephasing rate $\gamma$ depending linearly on temperature $T$. Technically, the detuning is achieved by increasing the temperature and varying the dependence of the cavity mode and trion energy on $T$. To model this, we expand to a linear order in the phonon strength of the detuning $2 \gamma \tau \approx \kappa\,\theta^2$, $\kappa \approx a_0 + a_1 (\omega_0 - E_t) / \Gamma$.

\section{Fitting of experimental data}
\noindent
The procedure for fitting the experimental data presented in Fig. 3 is as follows:

(i) First, from the Rabi oscillations (Fig. 3, main text) we derive the frequency correction $\tilde{\theta}$ extended by a second-order term $\tilde{\theta} = \theta + 0.0027 \,\theta^2$ and the phonon-induced decay $2 \gamma \tau = 0.0031\,\theta^2$.

(ii) We then model the $g^{(2)}(0)$ dependence as a function of the pulse area to determine the coupling strength $\delta = 0.72$.

(iii) Next, we correct the dependence of the trion time on the detuning and obtain the cavity mode width $\Gamma = 70~\mu eV$.

(iv) Finally, we plot $g^{(2)}(0)$ at $\theta = 2\pi$ against the detuning, using $a_1 = 0.014$.
\newline


The fit of $g^{(2)}(0)$ as a function of pulse length in Fig. 1 was performed using the full model including re-excitation and resonant scattering, as described in the main text, with a rectangular excitation  pulse. The parameters used are: $\Gamma = 85~\mu eV$, $g = 28~\mu eV$, $\gamma_{phon} = (0.1\,ps)\, \Omega^2_R$.

\bibliography{main}